\documentclass[10pt,conference]{IEEEtran}
\IEEEoverridecommandlockouts

\usepackage{cite}
\usepackage{bm}
\usepackage{amsmath}
\usepackage{extarrows}
\usepackage{amssymb}
\usepackage{graphicx}
\usepackage{color}
\usepackage{enumerate}
\usepackage{bookmark} 
\graphicspath{{figure}}
\usepackage{setspace}
\usepackage{subfigure}
\usepackage{algorithmic}
\usepackage{algorithm}
\usepackage{multirow}
\usepackage{stfloats}
\usepackage{nicematrix}
\usepackage{pifont}
\usepackage{threeparttable}
\usepackage{verbatim}

\usepackage[letterpaper,
top=0.85in,        % 顶部边距（要求≥0.7英寸）
bottom=0.85in,     % 底部边距
left=0.65in,       % 左边距
right=0.65in,      % 右边距（要求≥0.55英寸）
columnsep=0.25in   % 栏间距（建议≥0.24英寸）
]{geometry}

\newcommand{\BE}{\begin{equation}}
	\newcommand{\EE}{\end{equation}}
\newcommand{\BS}{\begin{subequations}}
	\newcommand{\ES}{\end{subequations}}
\renewcommand{\bf}{\bm}

\allowdisplaybreaks \allowdisplaybreaks[2]

\def\BibTeX{{\rm B\kern-.05em{\sc i\kern-.025em b}\kern-.08em
		T\kern-.1667em\lower.7ex\hbox{E}\kern-.125emX}}
\begin{document}

\title{Efficient Beamforming for Discrete SIM-Aided Multiuser Systems Under Statistical CSI\\
\thanks{\scriptsize This work was supported in part by the National Natural Science Foundation of China under Grant 62071275.
	$^{*}$ Corresponding author.
	%	$^{\dag}$ These authors contributed equally to this work.
	$^{\dag}$ Equal contribution.}
}

\author{\IEEEauthorblockN{
		Yuhui Jiao$^{\S\dagger}$,
		Qian Zhang$^{\S\dagger}$, 
		Xuejun Cheng$^\S$, 
		Yunxiao Li$^\S$,
		Yufei Zhao$^\ddagger$,  
		Ju Liu$^{\S *}$, and
        Yong Liang Guan$^\ddagger$}
\IEEEauthorblockA{$^\S$School of Information Science and Engineering, Shandong University, Qingdao, China \\
$^\ddagger$School of EEE Nanyang Technological University, Singapore \\
Emails: \{yuhuijiao2024, qianzhang2021, chengxuejun, yunxiaoli\}@mail.sdu.edu.cn, \\yufei.zhao@ntu.edu.sg, juliu@sdu.edu.cn, EYLGuan@ntu.edu.sg}}

\maketitle

\begin{abstract}
Stacked Intelligent Metasurfaces (SIM) have emerged as a revolutionary architecture for next-generation wireless communications, offering wave-domain signal processing capabilities with significantly reduced hardware complexity compared to conventional systems. However, most existing SIM research assumes continuous phase shifts and perfect instantaneous channel state information (CSI), which are impractical due to hardware discrete phase shift constraints and prohibitive pilot overhead. This paper presents a joint power allocation and discrete phase shift optimization framework for SIM-aided multiuser multiple-input single-output(MISO) downlink systems under statistical CSI. We formulate the achievable sum rate maximization problem considering practical discrete phase constraints and derive a closed-form expression for the average achievable rate under statistical CSI. To tackle the resulting non-convex optimization problem, we decouple the problem by using the weighted minimum mean square error (WMMSE) algorithm and alternating optimization (AO). Subsequently, we utilize the Lagrangian multiplier method and alternating direction method of multipliers (ADMM) to obtain closed-form iterative solutions. Our simulations demonstrate that the proposed algorithm reduces computational complexity by a factor of 50 compared to semi-definite relaxation (SDR) methods, , while maintaining over 85\% of the continuous phase shift performance with only 1-bit quantization, highlighting its feasibility for low-cost hardware systems.

\end{abstract}


\section{Introduction}

With the rapid development of the sixth generation (6G) of mobile communication technologies, emerging applications such as extended reality, holographic communications, and digital twins require increasingly high performance metrics including high bandwidth, low latency, and high reliability \cite{b1,b2}. To meet these diverse performance requirements and accommodate heterogeneous application scenarios, researchers have developed novel antenna architectures and configurations.

Multiple-input multiple-output (MIMO) technology has emerged as one of the most transformative innovations in wireless communication systems over the past two decades \cite{b3}. By exploiting spatial beamforming capacities through multiple antenna elements at both the transmitter and the receiver, MIMO systems demonstrate exceptional capabilities in enhancing spectral efficiency, improving link reliability, and increasing system capacity without requiring additional bandwidth or transmission power \cite{b4,zhao2023holographic}. However, traditional MIMO architectures equip each antenna element with dedicated radio frequency (RF) chain components such as power amplifiers, low-noise amplifiers and analog-to-digital converters. When the number of antennas increases further, the conventional construction results in prohibitive hardware complexity, excessive power consumption, and substantial manufacturing costs \cite{b5,b6}. To solve the high complexity of massive RF chains in conventional systems, stacked intelligent metasurfaces (SIM) have emerged as a promising solution \cite{b11,b12}. SIM consists of multiple sequentially cascaded programmable transmissive metasurface layers, with each layer containing numerous low-cost passive meta-atoms capable of independently adjusting the phase and amplitude to manipulate electromagnetic wave propagation. Through appropriate configuration of phase shifts at each SIM layer, the system can achieve flexible electromagnetic wave modulation and efficient beamforming operations directly in the wave domain \cite{zq11063223}\cite{Zhang2025XL_RIS}. The low-cost meta-atoms significantly reduce hardware complexity and power consumption compared to active RF chains equipped with power amplifiers and analog-to-digital converters. Moreover, the wave-domain signal processing capability of SIM not only reduces computational burden but also enables broadband analog beamforming with unprecedented flexibility. Through joint optimization of multi-layer phase configurations, SIM can synthesize arbitrary radiation patterns and achieve advanced spatial multiplexing for multiple users, making it an attractive candidate technology for next-generation wireless networks.

Due to its powerful wave-domain signal processing capabilities, SIM has been extensively investigated in wireless communication systems. The SIM-aided multiuser downlink communications was investigated in \cite{b114}, demonstrating the potential of SIM in enhancing downlink transmission efficiency. In \cite{Rezvani} , Rezvani \textit{et al.} studied SIM-aided uplink communication scenarios and proposed an interior point method to maximize the sum rate while accounting for hardware impairments. Additionally, the transmit precoding and receiver combining problems in SIM-aided holographic multiple-input multiple-output  communications were addressed through projected gradient methods \cite{10534211}. Furthermore, recognizing the advantages of analog processing, \cite{10679332} proposed a novel approach utilizing SIM for fully-analog wideband beamforming, which optimizes performance across a wide frequency range while maintaining analog simplicity. Beyond conventional cellular architectures, the synergy between SIM and cell-free network architectures has been explored in \cite{10535263}, where SIM was employed at access points to achieve highly energy-efficient information transfer. Moreover, \cite{10865993} extended the SIM implementation from the conventional base station side setup to the intermediate space between the base station and users, enabling more adaptive environmental adjustments as required.

Despite extensive research efforts, existing studies have predominantly overlooked two critical issues that are inevitable in practical systems. First, due to practical hardware constraints, the phase shift of each meta-atom in every SIM layer is inherently discrete rather than continuous, leading to unavoidable phase quantization errors. However, the aforementioned works have largely neglected this fundamental characteristic by assuming idealized continuous phase shift models. This simplification fails to capture the performance degradation caused by discrete phase constraints in realistic SIM deployments, thereby limiting the practical applicability of existing solutions. 
Second, a common assumption underlying these studies is the availability of perfect channel state information (CSI). However, acquiring accurate CSI in SIM-aided systems poses significant challenges. Due to the large number of meta-atoms across multiple SIM layers, obtaining perfect instantaneous CSI requires prohibitively high pilot overhead and frequent channel estimation, which consumes substantial bandwidth and computational resources. In practical deployments, the transmitter typically has access to only imperfect CSI. Furthermore, considering the enormous optimization space and computational complexity brought by the multi-layer structure of SIM, operating under statistical CSI becomes more practical, where the SIM meta-atom mode reconfiguration can be performed over relatively long time intervals and phase shift updates occur infrequently to reduce reconfiguration overhead.

To solve these issues, we propose a joint power allocation and discrete phase shift optimization method for SIM-based wireless communication system under statistical CSI. First, we derive a closed-form surrogate expression of the average achievable rate. For this non-convex problem, we employ the weighted minimum mean square error (WMMSE) algorithm and alternating optimization (AO) to decouple it into two subproblems\cite{b17}. Subsequently, we utilize the Lagrangian multiplier method and alternating direction method of multipliers (ADMM) to obtain closed-form iterative solutions\cite{b18}. Simulation results demonstrate that under large-scale SIM configurations, the proposed algorithm reduces computational runtime by 50 times compared to semi-definite relaxation (SDR) algorithms\cite{b19}.

\section{System Model}

\begin{figure}[t]	
	\centering \includegraphics[width=\linewidth]{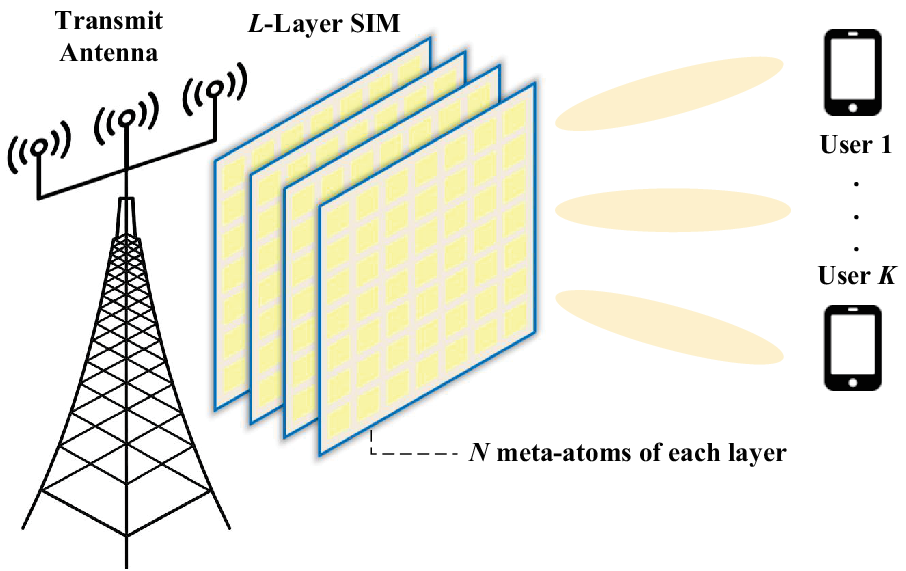}
	\vspace{-0.6cm}
	\caption{SIM-aided downlink MISO communication system model.}
	\label{fig1}
	\vspace{-0.8cm}
\end{figure}

As illustrated in Fig. \ref{fig1}, we consider a downlink multiuser multiple-input single-output (MISO) communication system, where the SIM is constructed by employing stacked metasurface architecture. In this system, a base station (BS) equipped with $M$ antennas is integrated with SIM at the transmitter to simultaneously serve $K$ single-antenna users. The SIM consists of $L$ metasurface layers, where each layer comprises $N$ independently controllable meta-atoms, thereby enabling direct precoding of the transmitted signals in the wave domain. For ease of analysis, we define the index sets for the users, metasurface layers, and meta-atoms per layer as $\mathcal{K} = \{1,\ldots, K\}$, $\mathcal{L} = \{1,\ldots, L\}$, and $\mathcal{N} = \{1,\ldots, N\}$ respectively.

For the $b$-bit SIM, the $n$-th meta-atom phase shift on layer \\[1pt]$l$ is denoted as $\phi_n^l = e^{j\theta_n^l}$, where $\theta_n^l \in \{0, \frac{2\pi}{2^b}, \dots, \frac{(2^b - 1)2\pi}{2^b}\}$, $n \in \mathcal{N}$, $l \in \mathcal{L}$. We denote the diagonal phase shift matrix of the $l$-th layer as $\boldsymbol{\Phi}^l = \text{diag}(\boldsymbol{\phi}^l) \in \mathbb{C}^{N \times N}$, where $\boldsymbol{\phi}^l = [\phi_1^l,\ldots,\phi_N^l]^T \in \mathbb{C}^{N \times 1}$. Additionally, $\mathbf{W}^l \in \mathbb{C}^{N \times N}$ ,  for any $l$ denotes the  attenuation coefficient matrix from the $(l - 1)$-th layer to the $l$-th layer. 
Based on the Rayleigh-Sommerfeld diffraction theory, the $(n, n')$-th entry $w^l_{n,n'}$ of  coefficient matrix $\mathbf{W}^l$ is given by \cite{b20}
\begin{equation}
	w^l_{n,n'} = \frac{d_x d_y \cos\varphi^l_{n,n'}}{d^l_{n,n'}} \left( \frac{1}{2\pi d^l_{n,n'}} - j\frac{1}{\lambda} \right) e^{j2\pi d^l_{n,n'}/\lambda},
\end{equation}
where $d_x$ and $d_y$ represent the horizontal width and vertical height of the elemental atom respectively, $\lambda$ is the wavelength, $d^l_{n,n'}$ denotes the corresponding propagation distance between the $n$-th element of the $(l-1)$-th layer and the $n'$-th element of the $l$-th layer, $\varphi^l_{n,n'}$ represents the angle between the propagation direction and the normal direction of the $(l-1)$-th layer. Similarly, let $\mathbf{w}_k^1 \in \mathbb{C}^{N \times 1}$ denote the coefficient from the transmit antenna array to the first layer of SIM. Based on the Huygens-Fresnel principle, the wave propagation between adjacent metasurface layers involves superposition. The resulting forward signal path through the SIM is thus described by\cite{10158690}
\begin{equation}
	\mathbf{G} = \boldsymbol{\Phi}^L \mathbf{W}^L \cdots \boldsymbol{\Phi}^2 \mathbf{W}^2 \boldsymbol{\Phi}^1 \in \mathbb{C}^{N \times N}.
\end{equation}

In this work, we employ the correlated Rayleigh fading distribution model to characterize the channel between the last layer of SIM and user $k$, which can be described as
\begin{equation}
	\bm{h}_k \sim \mathcal{CN}(\mathbf{0}, \beta_k \mathbf{R}), \quad \forall k \in \mathcal{K},
\end{equation}
where $\beta_k$ is the path-loss and $\mathbf{R} \in \mathbb{C}^{N \times N}$ is the spatial correlation of each surface with $tr (\mathbf{R}) = N$. Considering an isotropic scattering environment, the $(n, n')$-th entry of $\mathbf{R}$ can be expressed by \cite{b21}

\begin{equation}
	\mathbf{R}_{n,n'} = \text{sinc}\left(\frac{2\|\mathbf{u}_n - \mathbf{u}_{n'}\|_2}{\lambda}\right), \quad n,n' \in \mathcal{N},
\end{equation}
where $\mathbf{u}_n = \left[0, \frac{\bmod(n-1,N_r)d_x}{2}, \frac{\lfloor(n-1)/N_r\rfloor d_y}{2}\right]^\mathrm{T}$, $n \in \mathcal{N}$ 

represents the distance between $n$ and $n'$, $N_r$ represents the number of elements per row.

In this SIM-based transmitter beamforming system, the received signal at the $k$-th user can be written as
\begin{equation}
	y_k = \bm{h}_k^\mathrm{H} \mathbf{G} \sum_{i=1}^{K} \mathbf{w}_i^1 {p_i} s_i + n_k, \quad \forall k \in \mathcal{K},
\end{equation}
where $p_i^2$ denotes the allocated power for the user $i$, which satisfies $\sum_{i=1}^{K} p_i^2 \leq P_{\rm max}$, $P_{\rm max}$ is the budget of total BS transmit power, $n_k \sim \mathcal{CN}(0, \sigma_k^2)$ represents the noise at user $k$, which follows a circularly symmetric complex Gaussian distribution with variance $\sigma_k^2$. And the transmitted signal symbols $s_i$ are assumed to be independent and identically distributed (i.i.d.) random variables with zero mean and unit variance.

To obtain a closed-form surrogate expression of the average achievable rate in multiuser MISO systems, we introduce the following Lemma:

\textit{Lemma 1:} If $X = \sum_{i=1}^{t_1} X_i$ and $Y = \sum_{j=1}^{t_2} Y_j$ are both sums of nonnegative random variables $X_i$ and $Y_j$, where $t_1$ and $t_2$ denote the number of summation terms respectively, then we get the following approximation

\begin{equation}
	\mathbb{E}\left[\log_2\left(1 + \frac{X}{Y}\right)\right] \approx \log_2\left(1 + \frac{\mathbb{E}\{X\}}{\mathbb{E}\{Y\}}\right),
\end{equation}
\begin{equation}
	d_{as} = \left|\mathbb{E}\left[\log_2\left(1 + \frac{X}{Y}\right)\right] - \log_2\left(1 + \frac{\mathbb{E}\{X\}}{\mathbb{E}\{Y\}}\right)\right|,\label{das}
\end{equation}

where $\gamma$ is the Euler constant.where $d_{as}$ will gradually decrease as the number of antennas increases, and satisfies $d_{as} \in [0, \gamma/\ln(2)]$, where $\gamma$ is the Euler constant. And the approximation in (\ref{das}) becomes increasingly accurate in the large-scale antenna array due to the channel hardening effect.

\vspace{-0.1cm}
\textit{Proof:} See \cite{b22} and \cite{b14}.
\vspace{-0.1cm}

As a result, the downlink multiuser sum rate can be expressed as
\begin{equation}
	R = \sum_{k=1}^{K} \log_2(1 + \gamma_k),
\end{equation}
where $\gamma_k$ represents the signal-to-interference-plus-noise ratio (SINR).

\textbf{Proposition 1:} Under statistical CSI, the achievable sum rate can be approximated as:
\begin{equation}
	R \!=\! \sum_{k=1}^{K}\!\log_2\!\left(1 \!+\! \frac{\mathrm{tr}\left[p_k^2(\mathbf{G}\mathbf{w}_k^1)^\mathrm{H}\mathbf{R}_k(\mathbf{G}\mathbf{w}_k^1)\right]}{\sum_{i=1,i \neq k}^{K}\mathrm{tr}\left[p_i^2(\mathbf{G}\mathbf{w}_i^1)^\mathrm{H}R_k(\mathbf{G}\mathbf{w}_i^1)\right] + \sigma_k^2}\right)\!\!,\label{9}
\end{equation}

\textit{Proof:} By using the \textit{Lemma 1}, we can derive that

\begin{equation}
	\begin{aligned}
		R &= \mathbb{E}\left[\sum_{k=1}^{K}\log_2\left(1 \!+\! \frac{|p_k(\mathbf{G}\mathbf{w}_k^1)^\mathrm{H}\bm{h}_k|^2}{\sum_{i=1,i \neq k}^{K}|p_i(\mathbf{G}\mathbf{w}_i^1)^\mathrm{H}\bm{h}_k|^2 + \sigma_k^2}\right)\right] \\
		&\approx \sum_{k=1}^{K}\log_2\left(1 + \frac{\mathbb{E}\left[|p_k(\mathbf{G}\mathbf{w}_k^1)^\mathrm{H}\bm{h}_k|^2\right]}{\sum_{i=1,i \neq k}^{K}\mathbb{E}\left[|p_i(\mathbf{G}\mathbf{w}_i^1)^\mathrm{H}\bm{h}_k|^2\right] + \sigma_k^2}\right) \\
		&=\! \sum_{k=1}^{K}\!\log_2\!\!\left(1 \!+\! \frac{\mathrm{tr}\left[p_k^2(\mathbf{G}\mathbf{w}_k^1)(\mathbf{G}\mathbf{w}_k^1)^\mathrm{H}\mathbf{R}_k\right]}{\sum_{i=1,i \neq k}^{K}\mathrm{tr}\left[p_i^2(\mathbf{G}\mathbf{w}_i^1)^\mathrm{H}\mathbf{R}_k(\mathbf{G}\mathbf{w}_i^1)\right] \!+\! \sigma_k^2}\right)\!\!.
	\end{aligned}
\end{equation}

\vspace{-0.1cm}
\section{Beamforming Optimization}

In this section, based on the classical formulation in \cite{10158690}, we focus on the closed-form expression of sum rate in the discrete phase shift SIM-aided downlink multiuser MISO system.

The maximization problem is formulated as
\begin{subequations}\label{11}
	\begin{align}
		\quad \max_{\boldsymbol{\phi}^l, \mathbf{p}}& \quad f(\boldsymbol{\phi}^l, \mathbf{p}) = \sum_{k=1}^{K} \log_2\left(1 + \frac{S_k}{I_k}\right) \label{11a}\\
		\text{s.t} \quad & \mathcal{C}_G: \mathbf{G} = \boldsymbol{\Phi}^L\mathbf{W}^L \ldots \boldsymbol{\Phi}^2\mathbf{W}^2\boldsymbol{\Phi}^1, \label{11b}\\
		& \begin{aligned}\mathcal{C}_{\Phi}:  & \boldsymbol{\Phi}^l = \text{diag}\left(\boldsymbol{\phi}^l\right) = \text{diag}\left[\phi_1^l, \ldots, \phi_N^l\right], \\&\phi_n^l = e^{j\theta^l_n} , n \in \mathcal{N}, l \in \mathcal{L},\end{aligned} \\
		& \mathcal{C}_{\theta}:  \theta_{n}^{l} \in \left\{ 0, \frac{2\pi}{2^b}, \dots, \frac{(2^b - 1)2\pi}{2^b} \right\}, l \in \mathcal{L},\\
		& \mathcal{C}_p: \sum_{k=1}^{K} p_k^2 \leq P_{\rm max}, \\
		& \mathcal{C}_0: p_k \geq 0, \quad \forall k \in \mathcal{K}, 
	\end{align}
\end{subequations}
where $S_k$ and $I_k$ are the numerator and denominator of $\gamma_k$ obtained in \eqref{9}. Note that the constraint $\mathcal{C}_G$ expresses cascading beamforming in the wave domain, while $\mathcal{C}_{\Phi}$ and $\mathcal{C}_{\phi}$ indicate the discrete phase shift constraint, and $\mathcal{C}_p$ and $\mathcal{C}_0$ denote the BS maximum power constraint.

The discrete phase shift and its dependence on the unit manifold constraint make the problem tricky to solve. Additionally, in contrast to the single-layer optimization approaches employed for RIS, the multi-layer architecture of SIM introduces significant inter-layer coupling effects that substantially complicate the optimization process. Furthermore, this multi-layer configuration results in the computational complexity of traditional single-layer RIS optimization algorithms scaling unfavorably with the number of layers, thereby imposing considerable challenges on the computational efficiency and time requirements.

To solve this problem, we use WMMSE to convert the non-convex problem into a form that reduces computational complexity in SIM multilayer structures. Then use alternating optimization to iteratively optimize phase shift and power separately.

First, we use WMMSE method to transform the problem \eqref{11} into
\begin{equation}
	\begin{aligned}
		\min_{ \{\boldsymbol{\phi}^l \}^L_{l=1}, \boldsymbol{p}, \boldsymbol{u}_w, \boldsymbol{\rho}_w} &g(\boldsymbol{\phi}^l, \boldsymbol{p}, \boldsymbol{u}_w, \boldsymbol{\rho}_w) \quad \\ \text{s.t.} \quad \mathcal{C}_G, & \mathcal{C}_{\Phi}, \mathcal{C}_{\phi}, \mathcal{C}_p, \mathcal{C}_0,
	\end{aligned}
\end{equation}
where $\boldsymbol{\rho}_w \in \mathbb{R}_{++}^{K \times 1}, \boldsymbol{u}_w \in \mathbb{C}^{K \times 1},  \boldsymbol{p} = [p_1, p_2...,p_k]$, and
\begin{equation}
	\begin{aligned}
    	\mathcal{F} = &\sum_{k=1}^K \rho_{w,k} \left[ \sum_{i=1}^K (\mathbf{G}\mathbf{w}_i^{\mathrm{1}}p_i)^{\mathrm{H}} \mathbf{R}_k(\mathbf{G}\mathbf{w}_i^1p_i) + \sigma^2  \|u_{w,k}\|_2 \right. \\
    	&\left.- 2\mathbf{Re}\{u_{w,k}^{\mathrm{H}} \mathbf{R}_k^{\frac{1}{2}} (\mathbf{G}{\mathbf{w}_k^{1}}p_k)\}+1\right] - \log_2(\rho_{w,k}).
	\end{aligned}
\end{equation}

Then, we use the alternating optimization to solve the problem, the followings are the updates rules
\begin{subequations}
	\begin{align}
		\boldsymbol{u}_w^{\ell+1} &= \underset{\boldsymbol{u}_w}{\operatorname{argmin}} \; g(\boldsymbol{\phi}^{l,\ell}, \boldsymbol{p}^{\ell}, \boldsymbol{u}_w, \boldsymbol{\rho}^{\ell}_w), \label{14a}\\
		\boldsymbol{\rho}^{\ell+1} &= \underset{\boldsymbol{\rho}}{\operatorname{argmin}} \; g(\boldsymbol{\phi}^{l,\ell}, \boldsymbol{p}^{\ell}, \boldsymbol{u}^{\ell+1}_w, \boldsymbol{\rho}_w), \label{14b}\\
		\boldsymbol{p}^{\ell+1} &= \underset{\boldsymbol{p} \in \mathcal{C}_p \cap \mathcal{C}_0}{\operatorname{argmin}} \; g(\boldsymbol{\phi}^{l,\ell}, \boldsymbol{p}, \boldsymbol{u}^{\ell+1}_w, \boldsymbol{\rho}^{\ell+1}_w), \label{14c}\\
		\boldsymbol{\phi}^{l,\ell+1} &= \underset{\boldsymbol{\phi}^l \in \mathcal{C}_G \cap \mathcal{C}_{\bar{\Phi}} \cap \mathcal{C}_{\bar{\phi}}}{\operatorname{argmin}} \; g(\boldsymbol{\phi}^{l,\ell}, \boldsymbol{p}^{\ell+1}, \boldsymbol{u}^{\ell+1}_w, \boldsymbol{\rho}^{\ell+1}_w). \label{14d}
	\end{align}
\end{subequations}

By solving subproblems \eqref{14a} and \eqref{14b}, we can obtain the optimal closed-form solution of $u_w$ and $\rho_w$ as
\begin{align}
	u_{w,k}^* =& \frac{\mathbf{R}_k^{\frac{1}{2}} \mathbf{G} \mathbf{w}_k^{\mathrm{1}} p_k}{\sum_{i=1}^{K} p_i^2(\mathbf{G}\mathbf{w}_i^1)^\mathrm{H}\mathbf{R}_k(\mathbf{G}\mathbf{w}_i^1) + \sigma_k^2},\\
	\rho_{w,k}^* =& \left(1 - u_{w,k}^* \mathbf{R}_k^{\frac{1}{2}} \mathbf{G} \mathbf{w}_k^{\mathrm{1}} p_k\right)^{-1},
\end{align}
for any $k \in \mathcal{K}$.

\subsection{Optimize $\boldsymbol{p}$ With Fixed $\boldsymbol{\phi}^l$}
The Lagrangian function of the problem \eqref{14c} is given by
\begin{equation}
	\begin{aligned}
	    \mathcal{L}_p = &\sum_{k=1}^K \left[ (\mathbf{G}\mathbf{w}_i^{\mathrm{1}}p_i)^{\mathrm{H}} \mathbf{R}_k(\mathbf{G}\mathbf{w}_i^1p_i) + \sigma^2  \|u_{w,k}\|_2 \right. \\
	    &\left.- 2\mathbf{Re}\{u_{w,k}^{\mathrm{H}} \mathbf{R}_k^{\frac{1}{2}} (\mathbf{G}\mathbf{w}_k^{\mathrm{1}}p_k)\}+1\right] - \lambda(\sum_{i=1}^Kp_i-P_{\rm max}),
	\end{aligned}
\end{equation}
where $\lambda \geq 0$ denotes the dual variable.

Building upon the  established Lagrangian formulation, use duality theory to derive the optimal solution for power allocation. By taking the partial derivative of the Lagrangian function with respect to the power variable $p$ and setting it to zero, we obtain the first-order optimality conditions for the power optimization problem. Incorporating the Karush-Kuhn-Tucker (KKT) complementary slackness conditions, namely $\lambda \geq 0$, $p_k \geq 0$, and $\lambda_kp_k = 0$, we can derive the closed-form expression for the optimal transmit power of user $k$ as:
\begin{equation}
	p_k = \text{min}\left(P_{\rm max},\frac{\rho_{w,k}\mathbf{Re}\left\{u^\mathrm{H}_{w,k}\mathbf{R}^{\frac{1}{2}}_k\mathbf{G}\mathbf{w}^1_k\right\}}{\sum^K_{j=1}\rho_{w,j}{\mathbf{w}^{1,\mathrm{H}}_k}\mathbf{G}^\mathrm{H}\mathbf{R}_j\mathbf{G}\mathbf{w}^1_k\|u_{w,j}\|_2^2}\right).\label{18}
\end{equation}

Considering the increasing relationship between the Lagrange multiplier $\lambda$ and the total power constraint, use one-dimensional linear search to get the optimal solution within the feasible region. This computational framework demonstrates fast convergence properties and numerical efficiency.

\subsection{Optimize $\boldsymbol{\phi}^l$ With Fixed $\boldsymbol{p}$}

To avoid excessively large matrix dimensions during computation, we first transform $ \mathbf{G} \mathbf{w}_i^1 \sqrt{p_i}$ into
\begin{equation}
	\mathbf{G} \mathbf{w}_i^1 p_i = \mathbf{C}_i^l \boldsymbol{\phi}^l,
	\vspace{-0.2cm}
\end{equation}
\vspace{-0.2cm}
where
\begin{subequations}
	\begin{align}
		&\mathbf{C}_i^1 = \boldsymbol{\Phi}^L \mathbf{W}^L \cdots \boldsymbol{\Phi}^2 \mathbf{W}^2 \text{diag}(\mathbf{w}_i^1 p_i), \\
		&\mathbf{C}_i^L = \text{diag}(\mathbf{W}^L \cdots \boldsymbol{\Phi}^2 \mathbf{W}^2 \boldsymbol{\Phi}^1 \mathbf{w}_i^1 p_i), \\
		&\mathbf{C}_i^l = \boldsymbol{\Phi}^L \mathbf{W}^L \cdots \boldsymbol{\Phi}^{l+1} \mathbf{W}^{l+1} \text{diag}(\mathbf{W}^l \cdots \boldsymbol{\Phi}^1 \mathbf{w}_i^1 p_i).
	\end{align}
\end{subequations}

Then,  the problem \eqref{14d} with respect to $\boldsymbol{\phi}^l$ can be reformulated as 
\begin{equation}
	\min_{\boldsymbol{}{\phi}^l} \mathcal{G}(\boldsymbol{\phi}^l) = \boldsymbol{\phi}^{l,\mathrm{H}} \mathbf{B} \boldsymbol{\phi}^l - 2\mathrm{Re}\left\{\bm{d}^{\mathrm{H}} \boldsymbol{\phi}^l\right\}, \quad \text{s.t.} \quad \mathcal{C}_{\Phi}, \mathcal{C}_{\theta},\label{21}
\end{equation}
where $\mathbf{B} = \sum_{i=1}^K {\mathbf{C}_i^{l,\mathrm{H}}} \left[\sum_{k=1}^K \rho_{w,k} |u_{w,k}|^2 \mathbf{R}_k \right] \mathbf{C}_i^l$ and $\bm{d}^{\mathrm{H}} = \sum_{k=1}^K \rho_k u_k^* \mathbf{R}_k^{\frac{1}{2}} \mathbf{C}_k^l$.

However, the problem \eqref{21} remains intractable due to the non-convex constraint. Therefore, based on the ADMM principle, we introduce a new variable $\bm{x} = \boldsymbol{\phi}^l$, $l \in \mathcal{L}$ to decouple the complicated constraint as follows
\begin{equation}
	\begin{aligned}
		\min_{\boldsymbol{\phi}^l}& \quad \mathcal{G}(\boldsymbol{\phi}^l) = \boldsymbol{\phi}^{l,\mathrm{H}} \mathbf{B} \boldsymbol{\phi}^l - 2\mathrm{Re}\left\{\bm{d}^{\mathrm{H}} \boldsymbol{\phi}^l\right\} \quad\\ 
	    \text{s.t.}& \quad \mathcal{C}_{\bm{x}}:\bm{x} = \boldsymbol{\phi}^l, \bm{x}_n = e^{j\theta_n^l}, \theta_n^l \in \mathcal{C}_{\theta},
	\end{aligned}
\end{equation}

Then, we solve the problem by the following iterative updates
\begin{subequations}
	\begin{align}
		(\boldsymbol{\phi}^l)^{\ell+1} &= \underset{\boldsymbol{\phi}^l}{\arg\min} \, \mathcal{G}(\boldsymbol{\phi}^l) + \beta \|\boldsymbol{\phi}^l - \boldsymbol{x}^{\ell} - \boldsymbol{\omega}^{\ell}\|_2^2, \label{22a}\\
		\boldsymbol{x}^{\ell+1} &= \underset{\boldsymbol{x} \in \mathcal{C}_x}{\arg\min} \, \|\boldsymbol{x} - (\boldsymbol{\phi}^l)^{\ell+1} + \boldsymbol{\omega}^{\ell}\|_2^2, \label{22b}\\
		\boldsymbol{\omega}^{\ell+1} &= \boldsymbol{\omega}^{\ell} + \boldsymbol{x}^{\ell+1} - (\boldsymbol{\phi}^l)^{\ell+1}.\label{22c}
	\end{align}
\end{subequations}

The optimal solution for subproblem \eqref{22a} is given by the following
\begin{equation}
(\boldsymbol{\phi}^l)^{\ell+1} = (\mathbf{B} + \beta \mathbf{I}_N)^{-1} \left(\bm{d} + \beta (\boldsymbol{x}^l + \boldsymbol{\omega}^l)\right).
\end{equation}

Subproblem \eqref{22b} can be solved by projection onto the constant modulus constraint set $\mathcal{C}_{\bm{x}}$, which can be decomposed into $N$ independent scalar projections, where each element is projected onto the unit circle with discrete phase constraints:
\begin{equation}
	x_n^{\ell+1} = \mathcal{P}_{\mathcal{C}_{\boldsymbol{\theta}}} \left\{ \angle\left[({\phi}^l)_n^{\ell+1} - \omega_n^{\ell} \right] \right\},\label{25}
\end{equation}

	\begin{algorithm}[t]
	\caption{Proposed Algorithm For Problem \eqref{11}}
	\label{algorithm}
	\begin{algorithmic}[1]
		\STATE \textbf{Input:} Initialize $\mathbf{G}$, $\boldsymbol{\phi}^l$, $\bm p$, $\bm{u_w}$ and $\boldsymbol{\rho}_w$; Then, set the number of iteration $\ell = 0$.
        \REPEAT
        \STATE $u_{w,k}^{\ell+1} = \frac{\mathbf{R}_k^{\frac{1}{2}} \mathbf{G} \mathbf{w}_k^{\mathrm{1}} p_k}{\sum_{i=1}^{K} p_i^2 (\mathbf{G}\mathbf{w}_i^1)^\mathrm{H}\mathbf{R}_k(\mathbf{G}\mathbf{w}_i^1) + \sigma_k^2}, \quad k \in \mathcal{K}$;\\

        \STATE $\rho_{w,k}^{\ell+1} = \left(1 - u_{w,k}^* \mathbf{R}_k^{\frac{1}{2}} \mathbf{G} \mathbf{w}_k^{\mathrm{1}} p_k\right)^{-1}, \quad k \in \mathcal{K}$;
        \STATE \textbf{for} $k$ \textbf{from} $1$ \textbf{to} $K$\textbf{:}
        \STATE \hspace{0.5em} \textbf{repeat}
        \STATE \hspace{0.5em} \quad update $p^{\ell+1}_k$ according to \eqref{18};
        \STATE \hspace{0.5em} \textbf{until} $||f(({\boldsymbol{\phi}^l})^{\ell}, \mathbf{p}^{\ell}) - f(({\boldsymbol{\phi}^l})^{\ell}, \mathbf{p}^{\ell+1})|| < 10^{-5}$;
        \STATE \textbf{for} $l$ \textbf{from} $1$ \textbf{to} $L$\textbf{:}
        \STATE \hspace{0.5em} \textbf{repeat}
		\STATE \hspace{0.5em} \quad update $(\phi^l)^{\ell+1}$ according to \eqref{22a};
		\STATE \hspace{0.5em} \quad update $\bm{x}^{\ell+1}$ according to \eqref{22b};
        \STATE \hspace{0.5em} \quad update $\boldsymbol{\omega}^{\ell+1}$ according to \eqref{22c};
        \STATE \hspace{0.5em} \textbf{until} $||f(({\boldsymbol{\phi}^l})^{\ell}, \mathbf{p}^{\ell+1}) - f(({\boldsymbol{\phi}^l})^{\ell+1}, \mathbf{p}^{\ell+1})|| < 10^{-5}$;
		\STATE update $\mathbf{G}$ according to \eqref{11b};
		\UNTIL stopping criterion is satisfied.
	\end{algorithmic}
	\label{alg1}
\end{algorithm}
\vspace{-0.2cm}
where $\mathcal{P}_{\mathcal{C}_{\boldsymbol{\theta}}}\{\cdot\}$ denotes the projection operator that maps the phase to the nearest point in discrete phase set $\mathcal{C}_{\boldsymbol{\theta}}$. According to \cite{Zhang2025XL_RIS}, the closed-form solution of \eqref{25} can be derived as
\begin{equation}
	x_n^{\ell+1} = 
	\begin{cases}
		e^{j\frac{(\tau-1)\pi}{2^b - 1}}, & \text{if } \angle\left(  (	{\phi}^l_n)^{\ell+1} - {\omega^{\ell} _n}\right) < \frac{(2\tau - 1)\pi}{2^b}, \\
		e^{j\frac{\tau\pi}{2^b - 1}},     & \text{if } \angle\left(  (	{\phi}^l_n)^{\ell+1} - {\omega^{\ell} _n}\right) \geq \frac{(2\tau-1)\pi}{2^b},
	\end{cases}
\end{equation}

\noindent where $\tau = \left\lfloor \frac{2^{b-1} \angle\left(  (	{\phi}^l_n)^{\ell+1} - {\omega^{\ell} _n}\right) + \pi}{\pi} \right\rfloor.$
\vspace{0.1cm}

Finally, the overall algorithm is shown in \textbf{Algorithm 1}.

\section{Simulation Results}

In this section, we demonstrate the efficiency of the proposed algorithm in solving the SIM based achievable rate problem through simulation results. In simulations, we set the carrier frequency to 2 GHz, the length and width of meta-atoms are $\lambda$/2, the number of users and number of antennas $K = M = 5$, the BS power constraint $P_{\rm max}$ is 30 dBm, the noise power $\sigma^2$ is -80 dBm, the thickness of SIM $S_{\rm SIM}$ is 5$\lambda$ and the gap between layers is $S_{\rm SIM} / L$. The path loss coefficient $\beta_k = 10^{-3}d^{-2}_k$, where $d_k$ represents the distance from the last layer of SIM to user $k$. Additionally, users are uniformly and randomly distributed within an annulus centered at the base station, with an inner radius of 60 m and an outer radius of 80 m.

\begin{figure}[t]
	\centerline{\includegraphics[width=0.5\textwidth]{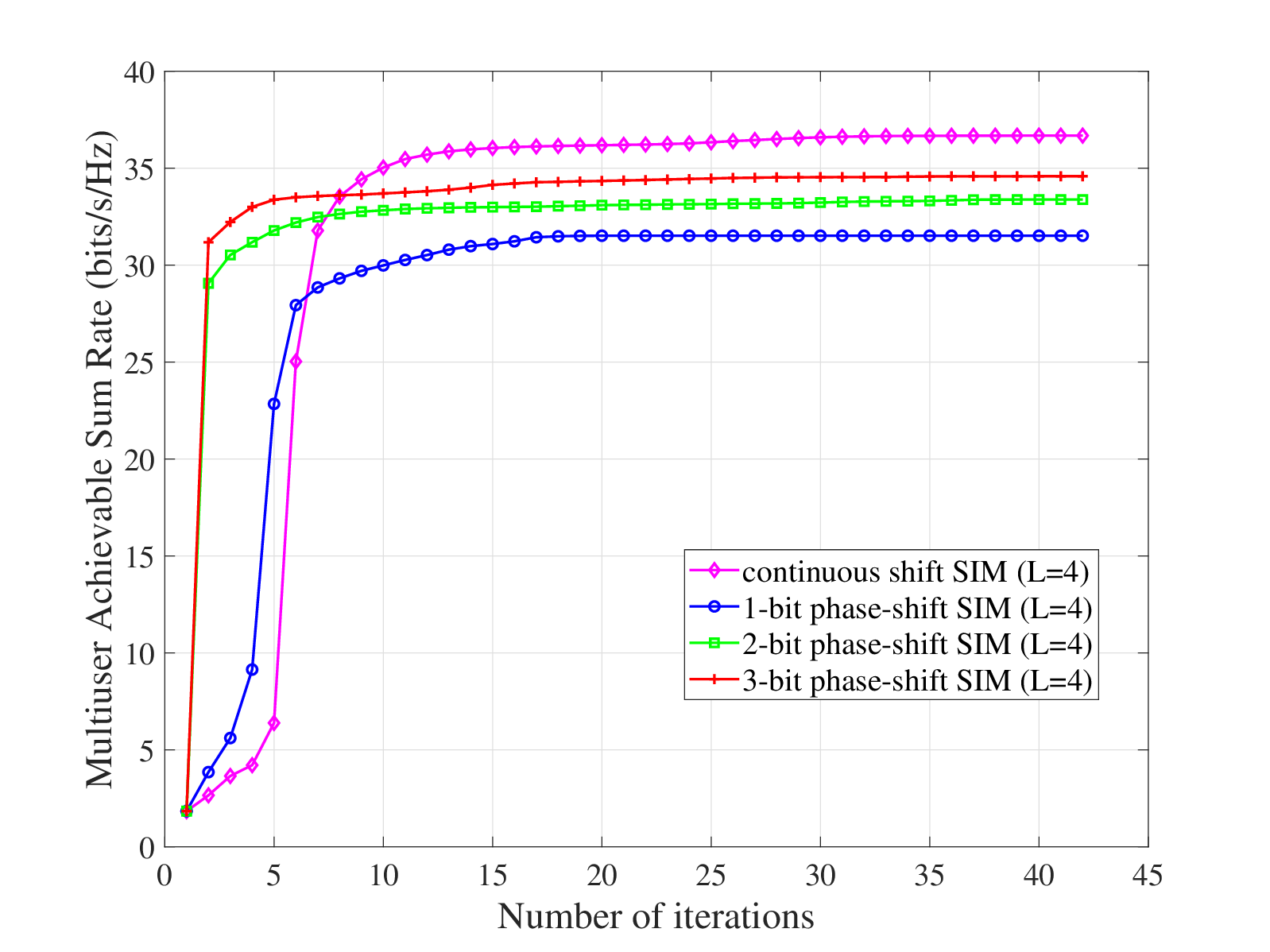}}
	\caption{Convergence performance of proposed algorithm in continuous and discrete phase shift SIM system.}
	\vspace{-0.5cm}
	\label{fig2}

\end{figure}

The convergence performance of the proposed algorithm for the SIM system is illustrated in Fig. \ref{fig2}, which demonstrates the downlink achievable sum spectral efficiency versus the number of iterations under different discrete phase shift resolutions. The simulation results reveal that all three configurations exhibit excellent convergence characteristics, reaching their steady-state values within 10 iterations. The result shows that even with 1-bit phase quantization, the SIM architecture can achieve over 85\% of the continuous phase-shift performance, supporting the feasibility of low-complexity hardware implementations for practical SIM systems.

\begin{figure}[t]
	\centerline{\includegraphics[width=0.5\textwidth]{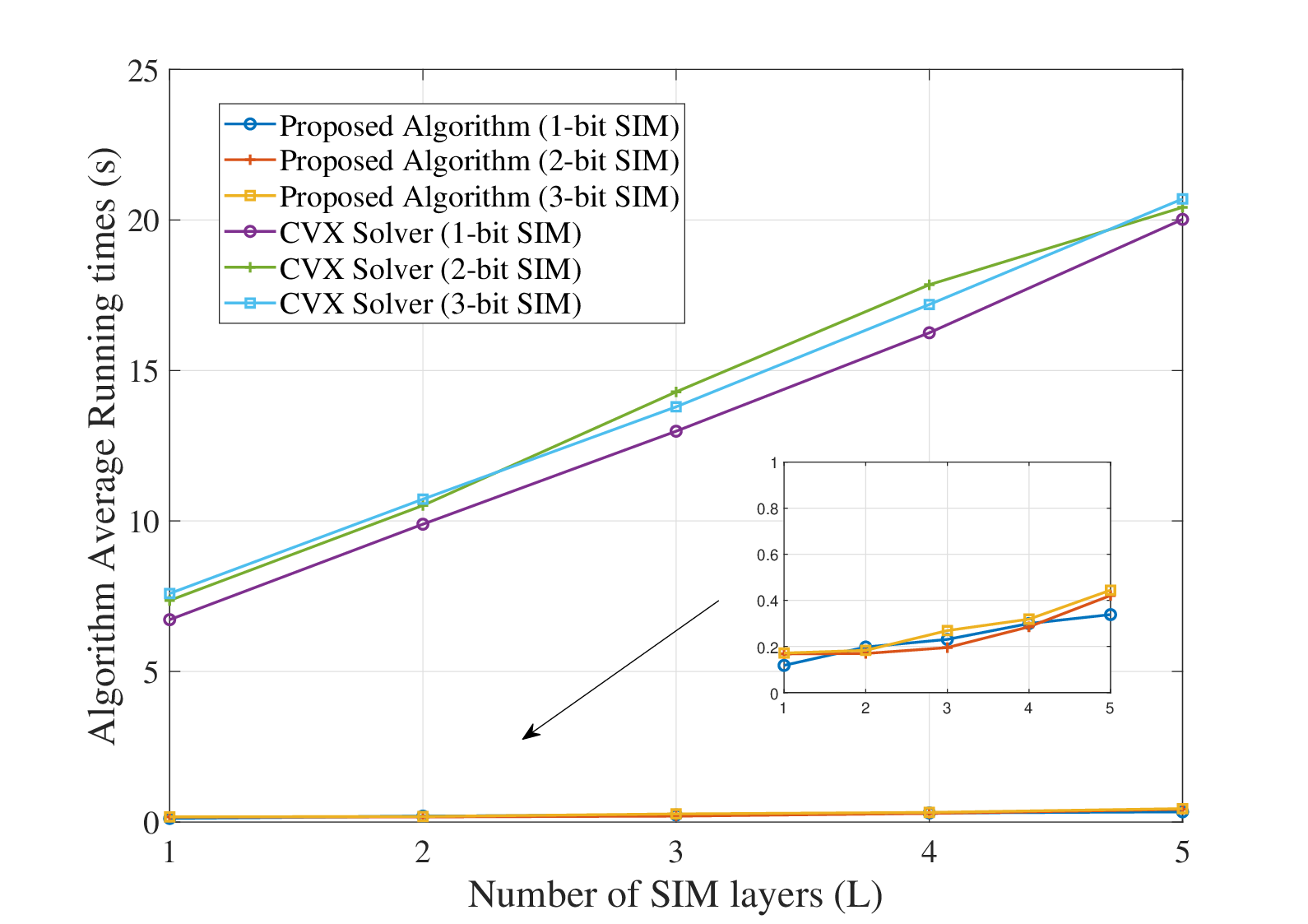}}
	\caption{Average algorithm runtime comparison with respect to the number of SIM layers.}
	\label{fig3}
	\vspace{-0.6cm}
\end{figure}

In Fig. \ref{fig3}, to further demonstrate the efficiency of the proposed algorithm, we compare the proposed algorithm and the SDR algorithms with CVX in terms of average running time. To reduce randomness, we ran 200 complete iterations and averaged the results for analysis. We can observe that the runtime of the SDR algorithm increases linearly with the number of layers, whereas our algorithm consistently maintains a low runtime regardless of the number of layers. Even in a single-layer SIM scenario, our algorithm achieves over ten times efficiency improvement. This provides a fast and effective solution for ultra-large-scale SIM applications.

\begin{figure}[t]
	\centerline{\includegraphics[width=0.5\textwidth]{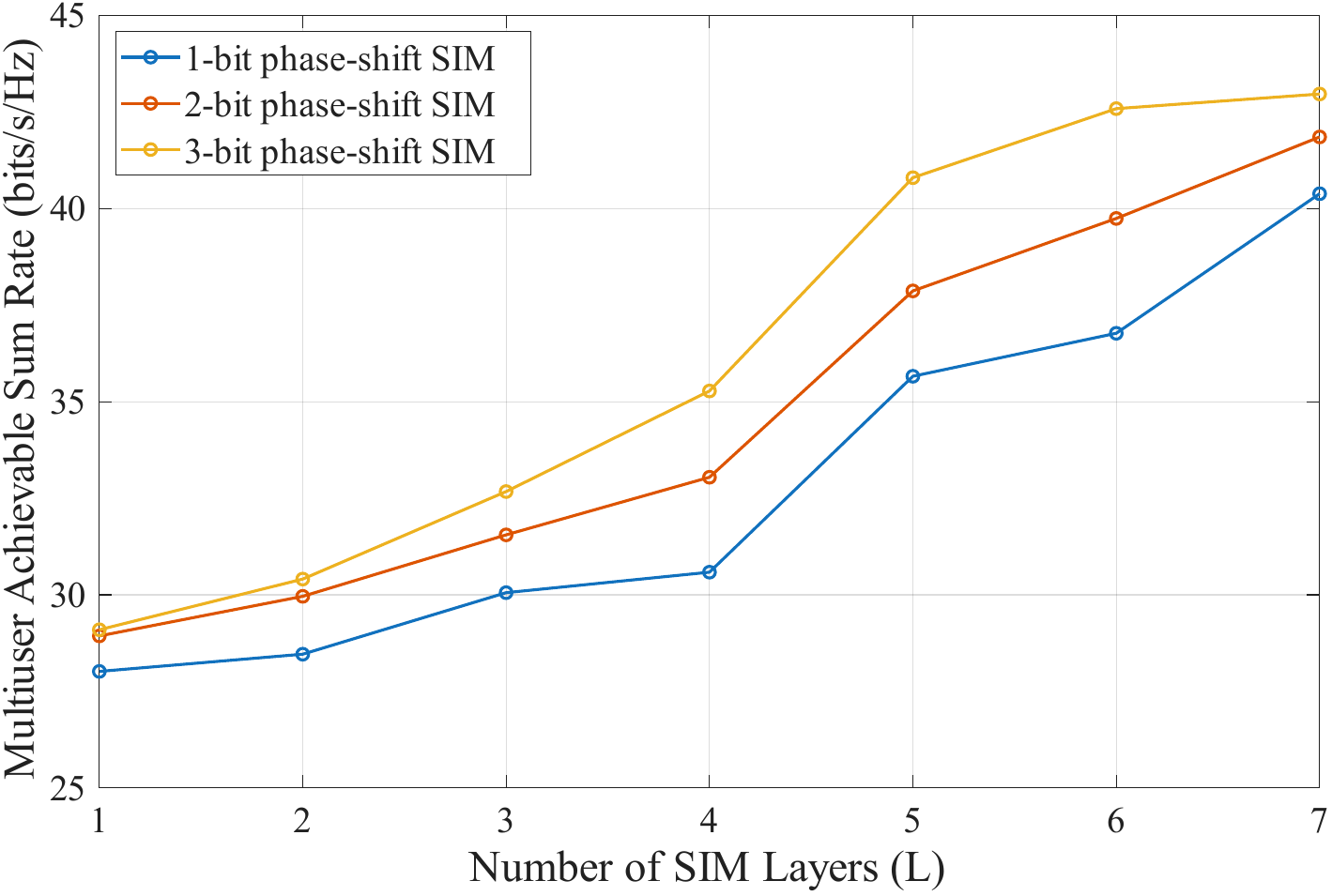}}
	\caption{Achievable sum rate of the large SIM-aided architecture with
		respect to the number of SIM layers.}
	\label{fig4}
\vspace{-0.5cm}
\end{figure}
Fig. \ref{fig4} shows that the achievable sum rate exhibits a monotonically increasing behavior with respect to the number of SIM layers across all quantization configurations.This enhancement can be attributed to the increased spatial degrees of freedom provided by additional SIM layers. Additionally, the 3-bit quantization scheme consistently outperforms the same layers of 2-bit and 1-bit schemes, achieving a maximum sum rate of approximately 43 bits/s/Hz at L = 7. While the 2-bit configuration demonstrates performance with approximately 4\% loss compared to 3-bit. These results suggest that 2-bit quantization may offer an attractive trade-off between hardware complexity and system performance. Additionally, results demonstrate the algorithm's scalability with respect to the number of SIM layers.

\vspace{-0.01cm}
\section{Conclusion}
This paper presents a comprehensive framework for joint power allocation and discrete phase shift optimization in SIM-assisted multiuser wireless systems. We derive closed-form expressions for achievable rates under statistical CSI and address the resulting non-convex optimization problem.The proposed method achieves significant computational complexity reduction compared to SDR-based approaches while demonstrating fast convergence and favorable scalability. Notably, even aggressive phase quantization (1-bit) retains substantial performance, validating the feasibility of low-complexity SIM architectures for next-generation wireless systems. Future work will extend this framework to uplink scenarios.

%%%%%%%%%% References %%%%%%%%%%%%%%
%\clearpage
\bibliographystyle{IEEEtran}
\bibliography{refs}

\end{document}